\documentclass[12pt]{article}

\newcommand{\rxte}{{\it RXTE}}
\newcommand{\xte}{{\it RXTE}}
\newcommand{\cxo}{{\it CXO}}
\newcommand{\chandra}{{\it Chandra}}

\newcommand{\tfe}{1E~1048.1$-$5937}
\newcommand{\tfn}{1E~2259$+$586}

\newcommand{\psr}{PSR~J1846$-$0258}

\usepackage{graphicx}
\usepackage{url}
\usepackage{color}

\usepackage{scicite}

\usepackage{times}

\newenvironment{sciabstract}{%
\begin{quote} \bf}
{\end{quote}}

\newcounter{lastnote}
\newenvironment{scilastnote}{%
\setcounter{lastnote}{\value{enumiv}}%
\addtocounter{lastnote}{+1}%
\begin{list}%
{\arabic{lastnote}.}
{\setlength{\leftmargin}{.22in}}
{\setlength{\labelsep}{.5em}}}
{\end{list}}

\title{Magnetar-like Emission from the Young Pulsar in Kes~75} 

\author
{F. P. Gavriil,$^{1,2,\ast}$ M. E. Gonzalez,$^{3}$ E. V. Gotthelf,$^{4}$  V. M. Kaspi,$^{3}$ \\ M. A. Livingstone,$^{3}$  and P. M. Woods$^{5,6}$\\
\\
\normalsize{$^{1}$NASA Goddard Space Flight Center, Code 662, Greenbelt, MD 20771,    USA}\\
\normalsize{$^{2}$CRESST, University of Maryland Baltimore County, Baltimore, MD 21250, USA}\\
\normalsize{$^{3}$Department of Physics, McGill University,}
\normalsize{3600 University St, Montreal, QC H3A 2T8, Canada}\\
\normalsize{$^{4}$Columbia Astrophysics Laboratory, Columbia University,}
\normalsize{New York, NY 10027-6001, USA}\\
\normalsize{$^{5}$Dynetics, Inc., 1000 Explorer Boulevard,    Huntsville, AL, 25806, USA}\\
\normalsize{$^{6}$NSSTC, 320 Sparkman    Drive, Huntsville, AL, 35805, USA} \\
\normalsize{$^\ast$To whom correspondence should be addressed; E-mail: gavriil@milkyway.gsfc.nasa.gov.}
}

\date{}

\begin{document} 


\baselineskip24pt


\maketitle



\begin{sciabstract}
We report detection of magnetar-like X-ray bursts from the young
pulsar \psr, at the center of the supernova remnant Kes~75.  This
pulsar, long thought to be rotation-powered, has an inferred surface
dipolar magnetic field of 4.9$\mathbf{\times}$10$\mathbf{^{13}}$~G,
higher than those of the vast majority of rotation-powered pulsars,
but lower than those of the $\mathbf{\sim}$12 previously identified
magnetars.  The bursts were accompanied by a sudden flux increase and
an unprecedented change in timing behavior. These phenomena lower the
magnetic and rotational thresholds associated with magnetar-like
behavior, and suggest that in neutron stars there exists a continuum
of magnetic activity that increases with inferred magnetic field
strength.
\end{sciabstract}


Magnetars are young, isolated neutron stars having ultra-high magnetic
fields\cite{td95,td96a}.  Observational manifestations of these exotic
objects include the Soft Gamma Repeaters (SGRs) and the Anomalous
X-ray Pulsars (AXPs).  Magnetars exhibit a variety of forms of
radiative variability unique to their source class; these include
short ($<$1~s) X-ray and gamma-ray bursts, and sudden flux
enhancements that decay on time scales of weeks to months, both of
which are too bright to be powered by rotational energy
loss\cite{wt06}.  A major puzzle in neutron star physics has been what
distinguishes magnetars from neutron stars that have comparably high
fields, yet no apparent magnetar-like emission\cite{km05}.

The 326-ms \psr\ is the central isolated neutron star associated with
the young shell-type supernova remnant (SNR) Kes~75 (SNR G29.6$+$0.1;
see ref~\citen{gvbt00} for details).  Assuming standard magnetic
dipole braking, this pulsar has among the largest dipolar magnetic
fields of the known young rotation-powered pulsars and the sixth
largest overall, $B\equiv
3.2$$\times$$10^{19}~\mathrm{G}~\sqrt{P\dot{P}} =
4.9$$\times$$10^{13}$~G, where $P$ is in seconds.  In addition, its
spin-down age of $\tau \equiv P/(n-1)\dot P = 884$~yr is the smallest
of all known pulsars\cite{gvbt00,lkgk06}.  The observed X-ray
luminosity of \psr\ is $L = 4.1$$\times$$10^{34}\left(d/
6~\mathrm{kpc} \right)^2$~erg~s$^{-1}$ in the 3$-$10~keV band, assuming
a distance of $d\sim6$~kpc, the mean distance found from HI and
$^{13}$CO spectral measurements\cite{lt08}.  The pulsar has all the
hallmarks of being rotation-powered -- a radiative output well under
its spin-down luminosity ($\dot E \equiv
3.9$$\times$$10^{46}\dot{P}/P^3~\mathrm{erg~s}^{-1} =
8.1$$\times$$10^{36}$~erg~s$^{-1}$), an otherwise unremarkable braking
index ($n =2.65$)\cite{lkgk06}, and a bright pulsar wind nebula (see
Fig.~1).
This pulsar is one of
only $\sim$3 young rotation-powered pulsars for which no radio
emission is detected, although this may be due to beaming.

Observations in the direction of Kes~75 obtained with the
\textit{Rossi X-ray Timing Explorer} (\rxte) have revealed several short
bursts of cosmic origin lasting $\sim$0.1~s (see Fig.~2).  We
discovered four bursts in a 3.4~ks observation made on 2006 May 31 and
a 5th in a 3.5~ks observation made on 2006 July 27.

These data were obtained with the Proportional Counter Array (PCA)
onboard \xte\ which provides $\sim$$\mu$s time resolution and 256
spectral channels over the $\sim$2$-$60~keV bandpass, and consists of
5 independent sub-units (PCUs). The bursts are plotted in Fig.~2 and
their properties are listed in Table~1.  We quantified the burst
properties as we have for those seen in bursting AXPs (see supporting
online text \citen{gkw02,gkw04,wkg+05,gkw06}).  All five bursts were
highly significant, and were recorded in all operational PCUs
simultaneously. We found no additional bursts in the 21.4~Msec of
available data of this field collected by \xte\ over the past 7 years.

Because of the PCA's large ($1^{\circ}$$\times$$1^{\circ}$)
field-of-view, the origin of the bursts was not immediately
apparent. However, we could unambiguously identify \psr\ as their
origin because the bursts coincided with a dramatic rise in its pulsed
flux, which lasted $\sim$2 months (see Fig.~2) and was remarkably
similar to those observed from AXPs\cite{kgw+03,ims+04,gk04}.  The
pulsed flux was extracted according to the method detailed in
ref~\citen{wkt+04} and corrected for collimator response and exposure
for each PCU.  We model the recovery from the pulsed flux enhancement
as an exponential decay (with $1/e$ time constant 55.5$\pm$5.7~day)
and estimate a total 2$-$60~keV energy release of
3.8$-$4.8$\times$$10^{41}\left(d/ 6~\mathrm{kpc} \right)^2$~erg,
assuming isotropic emission.  If we assume a power-law model for the
flux decay, commonly used for the magnetars, we obtain an index of
$-$0.63$\pm$0.06.  However, this model is rejected with
$\chi^2_{\nu}(51 \ \rm{DoF}) =1.31$ compared with $\chi^2_{\nu}(51
\ \rm{DoF})=0.95$ for the exponential model.

At the onset of the outburst, the timing noise of the source changed
dramatically from that typical of a young rotation-powered pulsar to
that typical of AXPs.  \psr\ was spinning down smoothly with a
braking index of $n=$2.65$\pm$0.01\cite{lkgk06} until phase coherence
was lost on MJD 53886, the same observation in which the first four
bursts were observed. This loss of phase coherence could signal a
spin-up glitch as has been seen to accompany other AXP radiative
events\cite{kgw+03,dkg07,tgd+07}. The dramatic sudden timing noise
makes the determination of accurate glitch parameters via
phase-coherent timing difficult. In the most recent data, the timing
noise appears to have settled somewhat, though has not relaxed to its
pre-burst behavior.

We also examined archival high-resolution CCD images of
Kes~75 obtained with the {\it Chandra X-ray Observatory} (\cxo) both
before (2000 Oct) and very fortuitously during (2006 June) the event. This
allows us to identify the dramatic change in the flux of the pulsar
relative to its bright, but relatively constant, pulsar wind nebula
(see Fig.~1 and supporting online text).

The \cxo-measured spectrum at the outburst epoch softened
significantly relative to quiescence. A fit to a power-law model in
2006 produced a larger value for the photon index, with
$\Gamma$$=$1.89$^{+0.04}_{-0.06}$ and 1.17$^{+0.15}_{-0.12}$ for epochs
2006 and 2000, respectively (3-$\sigma$ errors). Interestingly, the
larger value of the photon index is now closer to those seen in
magnetars ($\Gamma$$\sim$2$-$4). Due to this softening, the 0.5--2~keV
flux showed the largest increase, a factor of $17^{+11}_{-6}$, while
the 2$-$10~keV flux increased by a factor of $5.5^{+4.5}_{-2.7}$
(3-$\sigma$ errors, see Fig.~1). Though the 2006 spectrum is softer,
the large absorption precludes the identification of any significant
thermal components. Note that the \cxo\ spectral analysis was non-trivial due to
the brightness of the source and associated CCD pile-up; see online
supporting text for details.

The coincidence of the bursts with the flux enhancement (see Fig.~2),
the distinct changes in the pulsar spectral properties (see Fig.~1),
and the timing anomaly and sudden change in timing noise properties
all firmly establish \psr\ as the origin of the bursts.

This is the first detection of X-ray bursts from an apparent
rotation-powered pulsar. It is instructive to compare the burst
properties with those of SGRs and AXPs. SGRs are characterized by
their frequent, hyper-Eddington ($\sim$$10^{41}$~erg~s$^{-1}$), and
short ($\sim$0.1~s) repeat X-ray bursts.  AXPs also emit such bursts,
albeit less frequently\cite{gkw02}.  The bursts from \psr\ were short
($<$0.1~s), showed no emission lines in their spectra, and occurred
preferentially at pulse maximum.  The peak luminosities ($L_p$) of all
bursts were greater than the Eddington luminosity ($L_E$) for a
1.4~$M_{\odot}$ neutron star, assuming isotropic emission and a
distance of $d=6$~kpc\cite{lt08} (burst 2 had $L_p> 10 L_E$).
Considering the distribution of SGR and AXP burst temporal, energetic
and spectral properties\cite{gkw+01,gkw04}, the Kes~75 bursts are
indistinguishable from many of the bursts seen in AXPs and SGRs.

\psr's pulsed flux flare is also a magnetar hallmark.  A twisted
magnetosphere and associated magnetospheric currents induce enhanced
surface thermal X-ray emission, and resonant upscattering
thereof\cite{tlk02,bt07}. 
Flux enhancements and their subsequent decay in AXPs have been
interpreted as sudden releases of energy (either above or below the
crust) followed by thermal afterglow, in which case there is an abrupt
rise with a gradual decay.  A power-law fit was an excellent
characterization of AXP \tfn's flux decay after its 2002 outburst.
For \psr, such a model did not fit the data as well as an exponential
(see Fig.~2).
Spectral changes are also expected with these enhancements. The
softening of the source's spectrum suggests that it underwent a
transition from a purely magnetospheric-type spectrum, typical of
energetic rotation-powered pulsars, to one consistent with the
persistent emission from magnetars. For this reason, it is difficult
to directly compare the spectral characteristics of this flux
enhancement to those of other magnetars in outburst. The total
2$-$10~keV energy released during the flux enhancement
($3.3-3.8$$\times$$10^{41}\left(d/ 6~\mathrm{kpc} \right)^2$~erg,
assuming isotropic emission) is comparable that released in the 2007
flux enhancement \cite{tgd+07} of AXP
\tfe\ ($\sim$5$\times$$10^{42}\left(d/ 9~\mathrm{kpc} \right)^2$~erg),
the most most energetic enhancement yet seen from this AXP. It is also
comparable to the energy released during the rapid
($\sim$3$\times$$10^{39}\left(d/ 3~\mathrm{kpc} \right)^2$~erg) and
gradual ($\sim$2$\times$$10^{41}\left(d/ 3~\mathrm{kpc}
\right)^2$~erg) decay components of the 2002 outburst of AXP
\tfn\cite{wkt+04}. Similar to AXP \tfn's 2002 outburst \cite{wkt+04},
the energy released by \psr\ during the observed short bursts
represents only a small ($\sim$0.03\%) fraction of the total outburst
energy.

Prior to showing magnetar-like emission, \psr\ exhibited timing noise
and a glitch in 2001 \cite{lkgk06} that were both similar to what has
been seen observed in other comparably aged (i.e. $\tau$$\simeq$1~kyr)
rotation-powered pulsars. By contrast, in 2007, \psr\ exhibited much
larger timing noise, such that the root mean square phase residual
after subtracting a model including the spin frequency, and its first
and second derivative is a factor of $\sim$33 larger than before, for
the same duration of observations. Such a dramatic, sudden change in
timing noise characteristics has never been seen before in a
rotation-powered pulsar. The coincidence of the enhanced timing noise
with the flux flare is also reminiscent of behavior exhibited by AXP
\tfe\cite{gk04}.

Our discovery of distinctly magnetar-like behavior from what
previously seemed like a \textit{bona fide} rotation-powered pulsar
may shed new light on the magnetic evolution of these objects, and
whether their extreme fields originate from a dynamo operating in a
rapidly rotating progenitor\cite{td93a}, magnetic flux
conservation\cite{fw06}, or a strongly magnetized core, initially with
crustal shielding currents\cite{bs07}. In the first two scenarios,
magnetars are born with high magnetic fields which subsequently decay.
In the third recently proposed scenario, the very large magnetic
fields of magnetars slowly emerge as the shielding currents
decay\cite{bs07}. This source has a well measured braking index ($n
=2.65$$\pm$0.01)\cite{lkgk06}, at least before outburst, which is
significantly less than 3, suggesting that its spin properties, and
hence magnetic field are headed towards the magnetar
regime\cite{lyn04}. In this case, the timescale for magnetic field
decay, given by the magnetic field divided by its decay rate will be
$B/(\partial B/ \partial t$)$\sim$8~kyr, at which point \psr\ will
have $P\sim$1.3~s.  However, other mechanisms, such as the interaction
between a strong relativistic pulsar wind nebula (PWN) and the
magnetosphere\cite{hck99}, can also yield the value of $n$ measured
for \psr.  In this case, the magnetar-like behavior could be a result
of the moderately high $B$, with no $B$ evolution occurring.


There have been suggestions of magnetar-related emission from other
high-magnetic-field radio pulsars, e.g. PSR~J1119$-$6127\cite{gkc+05},
but, until now, nothing that could not also be explained within the
constraints of rotation-powered pulsar physics. It has been suggested
(see ref~\citen{km05}) that the high-$B$ field pulsars are related to
transient AXPs, magnetars generally in quiescence whose X-ray emission
can grow by factors of $\sim$hundreds in outburst.  Interestingly, the
first two reports of radio pulsations from a magnetar were from
transient AXPs after outburst\cite{crh+06,crh+07}.  Despite a lack of
radio emission, the behavior of \psr\ reinforces the connection
between transient AXPs and high-$B$ rotation-powered pulsars, and
suggests that careful monitoring of other high-$B$ rotation-powered
pulsars\cite{km05} is warranted.

The addition of \psr\ to the list of sources which emit magnetar-like
events provides insight into the origin of this activity.  Extreme
magnetic activity is prevalent in the SGRs which exhibit giant flares
with energy releases upwards of 10$^{44}$~ergs (see ref \citen{hbs+05}
for an example) and are also prolific busters, emitting bursts fairly
frequently, typically multiple times per year, with larger outbursts
occurring every few years. AXPs can be considered milder versions of
SGRs, with several showing sporadic short SGR-like events, though more
rarely than in SGRs, with even modest outbursts occurring only once or
twice per decade. Now, Kes~75, weakly magnetized by magnetar
standards, shows properties of both rotation powered pulsars and AXPs,
and seems to produce an outburst only roughly every decade.  The
detection of magnetar-like emission from Kes~75 suggests that there is
a continuum of ``magnetar-like'' activity throughout all neutron stars
which depends on spin-inferred magnetic field strength.




\begin{scilastnote}
\item We thank Tod Strohmayer for assistance and Alice Harding
and Demosthenes Kazanas for discussion.  MAL is a Natural Sciences and
Engineering Research Council (NSERC) PGS-D fellow.  Support for this
work was also provided by an NSERC Discovery Grant Rgpin 228738-03, an
R. Howard Webster Fellowship of the Canadian Institute for Advanced
Research, Les Fonds de la Recherche sur la Nature et les Technologies,
a Canada Research Chair and the Lorne Trottier Chair in Astrophysics
and Cosmology to VMK, and \xte\ grants NNG05GM87G and N5-RXTE05-34 to
EVG . This research has made use of data obtained through the High
Energy Astrophysics Science Archive Research Center Online Service,
provided by the NASA/Goddard Space Flight Center.
\end{scilastnote}



\clearpage
\includegraphics[width=\columnwidth]{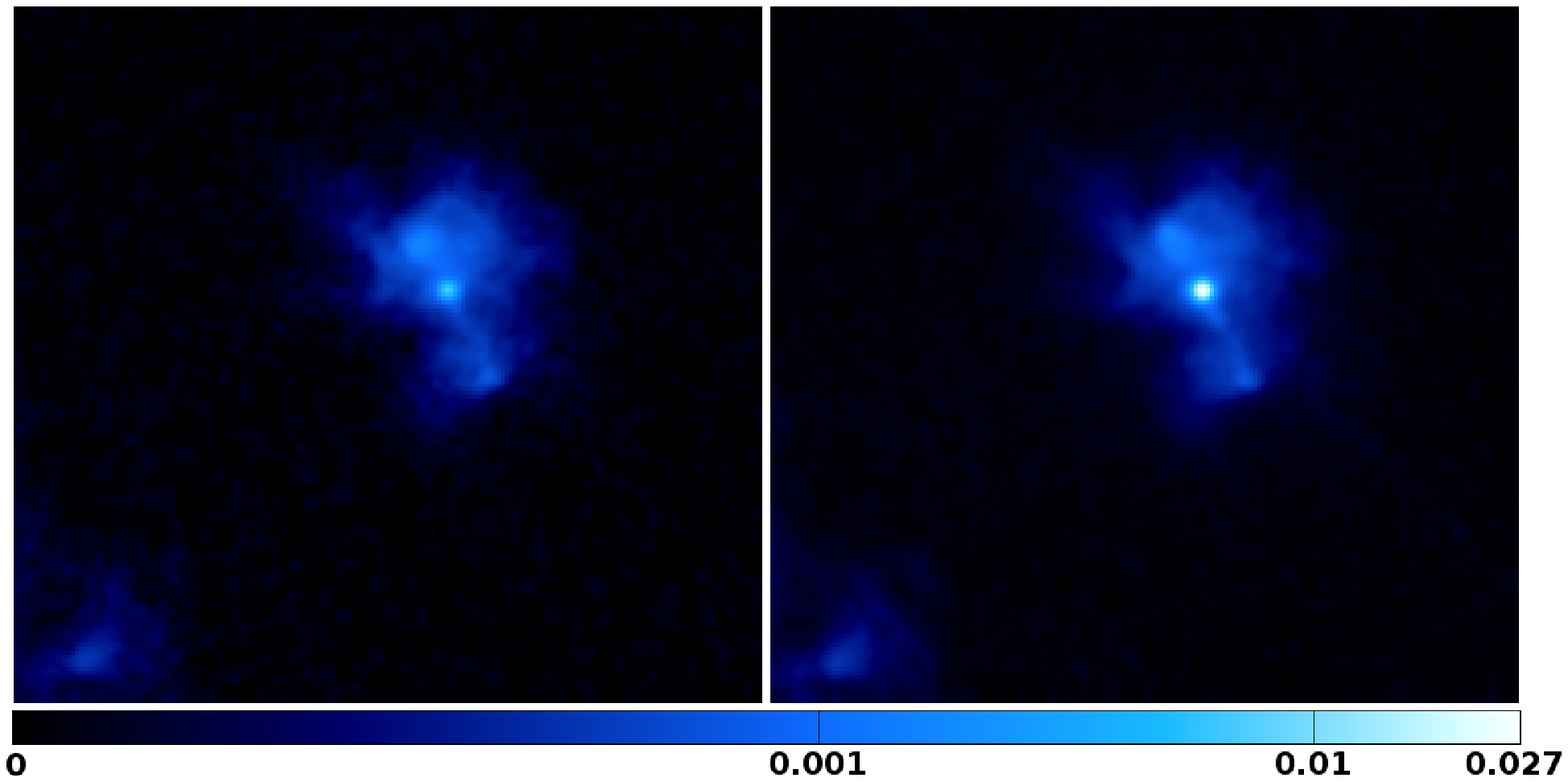}
\clearpage

\noindent {\bf Fig. 1.} High resolution \chandra\ X-ray images (0.5$-$10~keV) of
\psr\ in SNR Kes~75 centered on the pulsar and its surrounding PWN,
obtained before and during the 2006 outburst.  Following
the bursts, the pulsar became brighter as well as softer.  These
images were made using archival ACIS-S3 observations obtained on 2000
Oct 15-16 ({\it left}) and very fortuitously 2006 June 5, 7-8, 9, 12-13
({\it right}) and are background-subtracted, exposure-corrected,
smoothed with a constant Gaussian with width $\sigma$=0.5$''$ and
finally displayed using the same brightness scale.

\clearpage

\includegraphics[width=0.8\columnwidth]{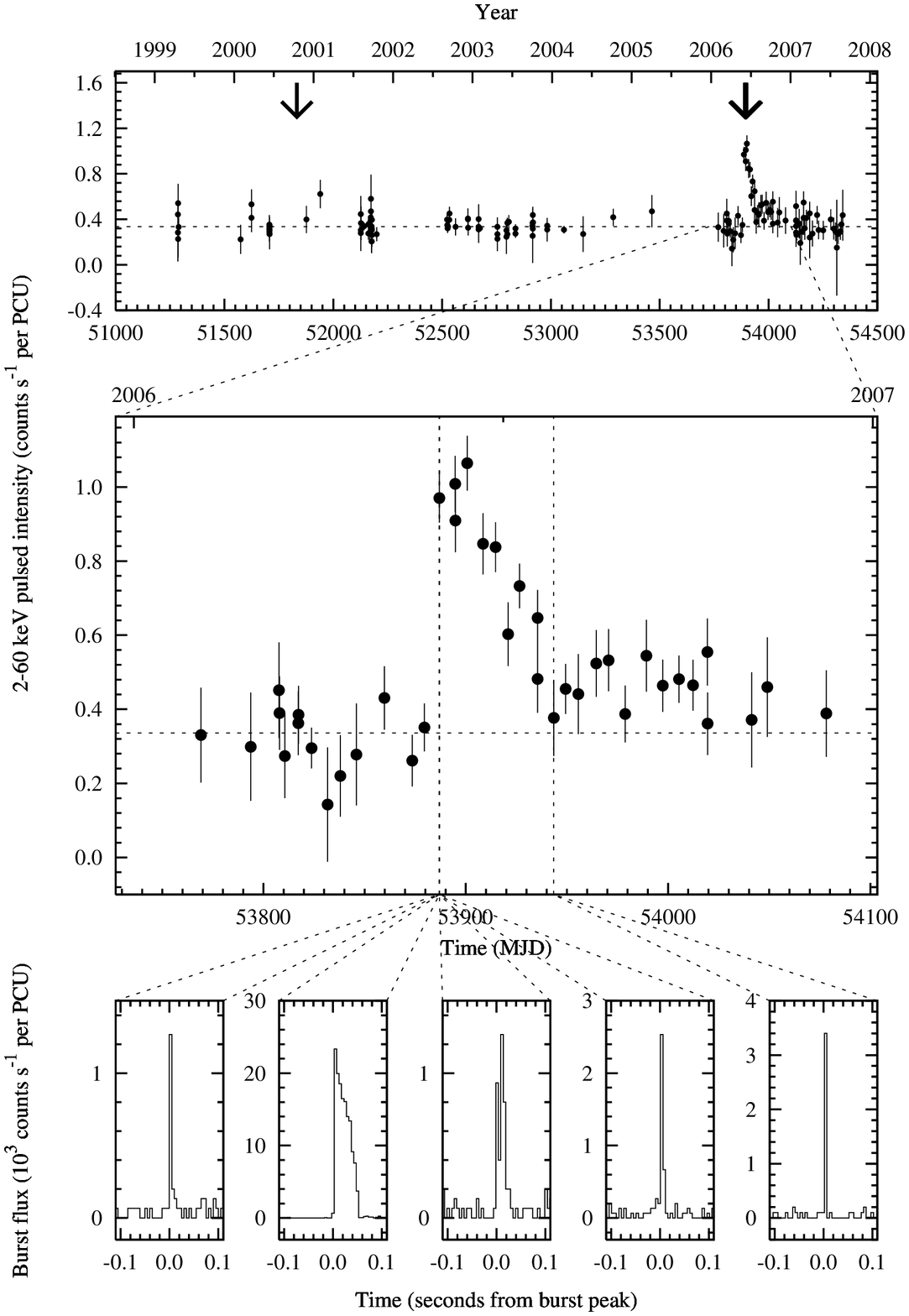}
\clearpage

\noindent {\bf Fig. 2.} Top: Pulsed flux history of \psr\ showing the
prominent outburst of June 2006 as recorded in the 2$-$60~keV band by
\rxte.  The horizontal dotted line represents the persistent flux
level. Epochs corresponding to \cxo\ observations are indicated with
arrows. Middle: The light curve around the outburst. The vertical
dashed lines indicate the epochs of the observations containing the
bursts, 2006 May 31 (4 bursts) and 2006 July 27 (1 burst). The
leftmost vertical dashed line also coincides with the time when phase
coherence was first lost. Bottom: The 2$-$60~keV \xte\ X-ray
lightcurves corresponding to five bursts detected from \psr, sampled
with 5~ms bins. The bursts lasted for $\sim$0.1~s and were detected
with high significance from two data sets obtained on 2006 May 31 and
2006 July 27.  Notice that in 7 years of \rxte\ observations the only
bursts found either occur at the onset of the $\sim$2 month X-ray
outburst (4 bursts) or at the end of the decay (1 burst).

\clearpage

\begin{center}
{
\begin{tabular}{lccccc}
\hline\hline
Table 1 & \multicolumn{5}{c}{\psr\ Burst Temporal and Spectral Properties}\\
\hline
      & \multicolumn{1}{c}{Burst 1} & \multicolumn{1}{c}{Burst 2} & \multicolumn{1}{c}{Burst 3} & \multicolumn{1}{c}{Burst 4} & \multicolumn{1}{c}{Burst 5} \\
\hline
\multicolumn{6}{l}{\textbf{Temporal properties}}\\
Burst day (MJD)                                  & 53886              & 53886                & 53886                & 53886                 & 53943   \\
Burst start time                               &  0.92113966(5)     & 0.93247134(1)       &  0.93908845(2)        &  0.94248467(5)      &  0.45543551(1) \\
(fraction of day) & & & & & \\
Rise time, $t_r$ (ms)                          & 4.2$^{+3.5}_{-2.0}$       & 1.1$^{+0.9}_{-0.5}$         & 1.90$^{+1.7}_{-0.9}$      & 4.1$^{+3.1}_{-1.9}$          & 0.9$^{+2.2}_{-0.7}$     \\
$T_{90}$ (ms)                          & 71.8$^{+38.0}_{-5.5}$ & 42.9$^{+0.3}_{-0.2}$ & 137.0$^{+11.4}_{-36.2}$ & 33.4$^{+29.1}_{-23.1}$ & 65.3$^{+0.7}_{-0.5}$  \\
Phase (cycles)                                 &   -0.49(1)                 &    -0.04(1)                   &        -0.20(1)             &       -0.05(1)                &  -0.08(1) \\
\multicolumn{6}{l}{\textbf{Fluences and fluxes}} \\
$T_{90}$ Fluence                  & 8.9$\pm$0.7 & 712.8$\pm$2.5 & 18.3$\pm$0.7 & 18.4$\pm$0.7 & 18.4$\pm$1.1 \\
(counts/PCU) &&&&&\\
$T_{90}$ Fluence   & 4.1$\pm$2.4 & 289.9$\pm$13.1 & 6.6$\pm$2.5 & 5.8$\pm$1.7 & 5.3$\pm$2.0  \\
(10$^{-10}$~erg/cm$^{2}$)  & & & & & \\
Flux for 64~ms                & 57$\pm$36 & 4533$\pm$227 & 99$\pm$41 & 97$\pm$31 & 79$\pm$32 \\
(10$^{-10}$ erg/s/cm$^{2}$) &&&&&\\
Flux for $t_r$              & 678$\pm$427 & 5783$\pm$885 & 810$\pm$385 & 828$\pm$284 & 2698$\pm$1193 \\
(10$^{-10}$ erg/s/cm$^{2}$)  &&&&&\\
\multicolumn{6}{l}{\textbf{Spectral properties}} \\
Power-law index             &  0.89$\pm$0.58 &  1.05$\pm$.04  & 1.14$\pm$0.34 & 1.36$\pm$0.25 & 1.41$\pm$0.31 \\
$\chi^2/$DoF (DoF)             & 0.42 (1) & 1.16 (55)  & 0.97 (3) & 0.35 (2) & 1.18 (2) \\
\hline
\end{tabular}}
\end{center}
\clearpage

\noindent {\bf Table 1.}  All the quoted errors represent 1-$\sigma$
uncertainties unless otherwise indicated.  All times are given in
units of UTC corrected to the Solar System barycenter using the source
position
R.A.=$18^{\mathrm{h}}46^{\mathrm{m}}24$\mbox{$.\!\!^{\mathrm{s}}$}94,
decl=$-02^{\circ}58^\prime30$\mbox{$.\!\!^{\prime\prime}$}1 and the
JPL DE200 ephemeris\cite{hcg03}.

\clearpage 

\section*{Supporting Online Text}

\paragraph*{Burst Properties.} 
We defined the burst peak time as the midpoint between the two events
having the shortest separation in the peak bin of the 31.25-ms
digitized 2$-$60~keV PCA lightcurve. The error on the burst peak is
determined using the rise time ($t_r$) method outline in
ref~\citen{gkw04}.  The burst background rates were measured by
averaging the adjacent 0.5-s flux.  We used a sliding 64~ms boxcar on
an event-by-event level to determine peak flux.  The total burst
fluence is calculated by integrating the events within a 0.1~s
interval centered on the burst peak and subtracting the modeled
background component. $T_{90}$ is defined as the time between when the
burst fluence goes from 5\% to 95\% of its total fluence. Burst
spectra were extracted from a 1.2~s interval in the lightcurve
centered on the peak emission as defined above.  The background is
estimated from the same adjacent interval as used for the burst
fluence and flux analysis.  Spectra were grouped for a minimum of 15
counts per bin after background subtraction and fitted with a absorbed
power-law model in the 2$-$60~keV energy range using \texttt{XSPEC}.
The column density was held fixed at $N_H$=4$\times$10$^{22}$
cm$^{-2}$, the value found by ref~\citen{hcg03} and
ref~\citen{msb+07}.  Response matrices were generated using the standard
software. This provided a good fit for burst~2 which had the most
counts, significantly better then for an absorbed blackbody model. The
statistics for the other bursts were too poor to distinguish
models. To calibrate the burst fluxes and fluences we calculated a
factor from the 2$-$60~keV count rate to power-law flux (unabsorbed) in
the same band using the burst's power-law index (see Table~1), and multiplied our
total fluence and peak fluxes by this factor.

\paragraph{Imaging Observations of \psr.}
The data were processed using the CIAO v3.3 and CALDB v3.2.2 software
and subjected to the standard processing, resulting in effective
exposure times of $\sim$37~ks and $\sim$154~ks for the 2000 and 2006
observations, respectively.  For the spectral analysis, great care is
needed as these two observation are strongly effected by CCD
``pile-up''\cite{dav01,url1}, where two or more photons are recorded
in a single CCD pixel, thus distorting the overall spectrum. Each
observation is uniquely effected because it depends on count rate and
the CCD read-out times (3.2~s and 1.8~s for the 2000 and 2006
observations, respectively).  The background-subtracted count rate for
the first epoch was $0.092\pm0.002$~cts~s$^{-1}$ in the 1--10 keV
range with a pile-up fraction of $6\pm4\%$. Despite a faster read-out
time for the 2006 observation, the pile-up fraction increased to
$25\pm3\%$ for the 2006 observations due to the higher pulsar flux of
$0.330 \pm 0.005$~cts~s$^{-1}$.  To take into account pile-up in our
spectral analysis, we followed the prescription of ref~\citen{dav01}.
Spectra from the pulsar were extracted from a $2^{\prime\prime}$
radius aperture and the background estimated from a
$2^{\prime\prime}<r<4^{\prime\prime}$ annular region, using a minimum
of 50 cts bin$^{-1}$. No significant spectral changes were detected
within the four 2006 observations and these spectra were fit
simultaneously to a piled-up power-law model with the absorbing column
fixed at $N_H$=4$\times$10$^{22}$ cm$^{-2}$.  This provides a best-fit
photon index, $\Gamma$, for the 2000 and 2006 observations of
1.17$^{+0.15}_{-0.12}$ and 1.89$^{+0.04}_{-0.06}$, respectively
(3$\sigma$ errors).  Despite the pile-up, the 2006 observations show a
clear softening of the spectrum.  In turn, the unabsorbed fluxes in
units of 10$^{-11}$~erg~s$^{-1}$~cm$^{-2}$ for the 2000 (2006)
observations were 0.10$^{+0.01}_{-0.01}$ (1.7$^{+0.8}_{-0.5}$) in the
0.5--2~keV range, and 0.42$^{+0.16}_{-0.05}$ (2.3$^{+1.4}_{-0.7}$) in
the 2$-$10~keV range (3$\sigma$ errors).  Due to the high count rate in
the 2006 observations, significant emission from the pulsar was
detected during the readout interval, resulting in a ``readout streak"
that contains un-piled, real events from the source.  The power-law
spectral parameters derived using these data are in agreement with
those listed above.

\end{document}